\documentclass[conference]{IEEEtran}
\IEEEoverridecommandlockouts 

\usepackage{cite}
\usepackage{upgreek}
\usepackage{amsfonts}

\usepackage{graphicx}

\usepackage{url}

\usepackage{gensymb}
\usepackage{amsmath}

\begin{document}
%
\title{Steady-State Thermal Analysis of an Integrated 160 GHz Balanced Quadrupler Based on Quasi-Vertical Schottky Diodes}

\author{
\IEEEauthorblockN{
Souheil Nadri\IEEEauthorrefmark{1},
Linli Xie\IEEEauthorrefmark{1}, Naser Alijabbari\IEEEauthorrefmark{1}, John T. Gaskins\IEEEauthorrefmark{2}, Brian M. Foley\IEEEauthorrefmark{2}, Patrick E. Hopkins\IEEEauthorrefmark{2} and \\ Robert M. Weikle II\IEEEauthorrefmark{1}
}
\IEEEauthorblockA{\IEEEauthorrefmark{1}Department of Electrical Engineering, University of Virginia, Charlottesville, VA 22904 U.S.A\\ Email: sn3t@virginia.edu\\\IEEEauthorrefmark{2} Department of Mechanical Engineering, University of Virginia, Charlottesville, VA 22904 U.S.A}
}


\maketitle

\begin{abstract}
This work reports on a steady-state thermal analysis of a 160 GHz balanced quadrupler, based on a quasi-vertical varactor Schottky diode process, for high power applications. The chip is analyzed by solving the heat equation via the 3D finite element method. Time-Domain Thermoreflectance (TDTR) was used to measure the thermal conductivity of the different materials used in the model. A maximum anode temperature of 64.9$\degree$C was found from the simulation. The addition of an extra beam lead connected to the block, for heat sinking, was found to reduce this maximum temperature to 41.0$\degree$C.
\end{abstract}

\IEEEpeerreviewmaketitle

\section{Introduction and Background}

\IEEEPARstart{S}{chottky} 
diode multiplier chain technology constitutes a very good candidate to deliver high power outputs up to 1 THz.
Progress in power amplifiers have made it possible to achieve watt levels of power in the W-band (75-110 GHz). Therefore, the input stage in the multiplier chain needs to be able to withstand large power levels, without any significant degradation in electrical performance. Consequently, thermal management considerations become very crucial in the design of frequency multipliers [1]. \newline
The quasi-vertical diode process developed at the University of Virginia [2] not only addresses series resistance concerns [2], but also thermal management by integrating the diodes onto a silicon substrate, utilizing the higher thermal conductivity of silicon compared to gallium arsenide. This paper presents a 3D finite element analysis of a 160 GHz balanced quadrupler based on varactor diodes that use this process. The thermal conductivities of the different materials comprising the multiplier are obtained using thermoreflectance measurements.  

\section{Quasi-Vertical Diode Geometry}
The quasi-vertical Schottky diode geometry, thoroughly discussed in [2][3], consists of a thin GaAs mesa (1.28  $\upmu $m) that has been wafer-bonded to a host silicon-on-insulator (SOI) substrate (Fig.1). The ohmic contact is formed prior to the epitaxial transfer, and the GaAs is not relied upon for structural integrity. The primary motivation for developing this diode structure was to reduce device parasitics, especially the series resistance [2]. However, the quasi-vertical structure is also advantageous for thermal handling, because it heterogeneously integrates GaAs Schottky diodes onto Si substrates, which have a higher thermal conductivity than GaAs (see Table I), thus providing a better heat sink. 

\begin{figure}[h]
\centering
\includegraphics[width=1.8 in]{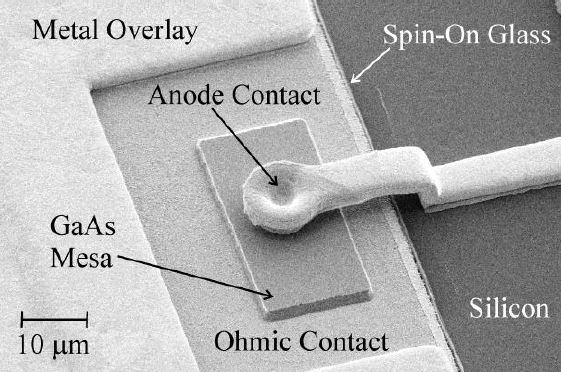}
\caption{Scanning electron micrograph of a quasi-vertical Schottky diode integrated on a 15 $\upmu$m silicon substrate.}
\label{fig_scattering}
\end{figure}
 
\section{Time-Domain Thermoreflectance Characterization}
Heat transport properties of materials, such as thermal conductivity, can vary significantly from bulk properties for micro and nano scale devices [4]. In fact, scattering mechanisms such as boundary, impurity, and imperfection scatterings, which are more pronounced in thin films, lead to a reduction in the thermal conductivity of materials [4]. In order to account for these phenomena in our simulated model for the 160 GHz quadrupler, time-domain thermoreflectance (TDTR) was used to measure the thermal properties of the material layers that make up this device.\\  The technique and data analysis is described in detail elsewhere [5][6]; as such, the technique is briefly described here.  TDTR is a non-contact optical pump-probe measurement, shown here in Figure 2, that utilizes a modulated short-pulsed heating event, the pump, to heat up the surface of a metal transducer, and a time-delayed low-power pulse, the probe, to quantify changes in thermal decay over time.  The pump pulse is modulated via an electro-optical modulator at a driving frequency of 8.8 MHz.  The probe pulse is delayed in time, relative to the pump heating event, by a mechanical delay stage at the same spatial point as the pump heating event.  An RF lock-in amplifier is used to detect changes in the reflected probe signal at the modulation frequency of the pump heating event.  The ratio of the in-phase to out-of-phase voltage (-Vin/Vout) yields information about the thermal properties of the underlying layers via well known relations [5][6].
\begin{figure}[h]
\centering
\includegraphics[width=2.8 in]{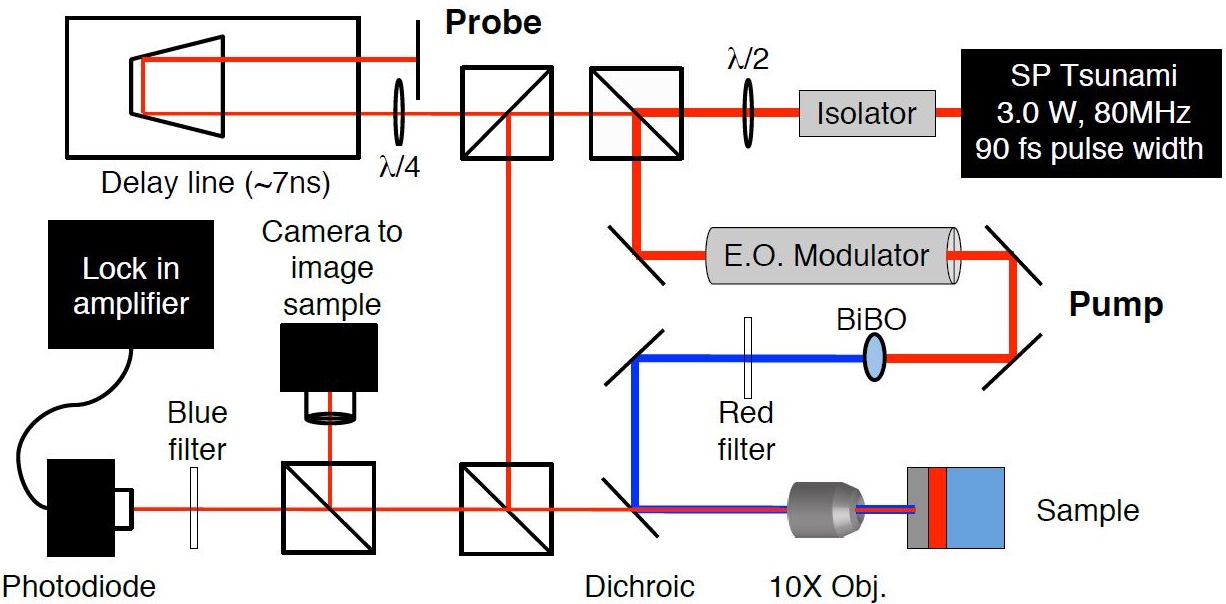}
\caption{TDTR measurement setup.}
\label{fig_scattering}
\end{figure}

Table I summarizes the thermal conductivity values, measured using TDTR, of the Si layer in the silicon-on-insulator (SOI) wafer, gallium arsenide, and the bonding agent spin-on-glass (SOG), used in the thermal analysis.

\begin{table}[h]
\caption{Measured thermal conductivities (${\rm W. m^{-1}. K^{-1}}$) used in the analysis.}
\label{table_example}
\centering
\begin{tabular}{|c|c|c|c|c|}
\hline
Material & Thickness ($\upmu$m) & Bulk 
 & Measured & Meas. Uncertainty

\\
\hline
Si in SOI & 15.00 & 150.00 & 115.00 & 12.4\%\\
\hline
GaAs &1.28 & 46.05 & 42.88 & 7.5\%\\
\hline
SOG & 0.50 & 0.60 & 0.22 & 10.0\%\\
\hline
\end{tabular}
\end{table}

In addition, the thermal conductivity of the ohmic contact was also characterized via the electrical resistivity (four-point probe) and the Wiedemann-Franz law. The specific alloy investigated in this work is a Ni(30 nm)-Ge(40nm)-Au(400nm) based alloy with a 10 nm Ti adhesion layer. Three options were investigated: annealed alloy with 50 nm evaporated gold and 350 nm plated gold (option 1), annealed alloy with 400 nm evaporated gold (option 2), and finally non-annealed alloy with both plated and evaporated gold. The results for the three ohmic contact options are summarized in Table II. 

\begin{table}[h]
\caption{Measured thermal conductivities (${\rm W. m^{-1}. K^{-1}}$) for the three ohmic contact options.}
\label{table_example}
\centering
\begin{tabular}{|c|c|c|c|}
\hline
Material  
 & Measured Value & Measurement Uncertainty

\\
\hline
Option 1 & 31.1 & 6\%\\
\hline
Option 2 & 42.2 & 7\%\\
\hline
Option 3 & 108.7 & 5\%\\
\hline
\end{tabular}
\end{table}

\section{Simulation Results and Discussion}
The chip analyzed in this work is based on the 160 GHz balanced quadrupler detailed in [3]. The finite element method solution to the heat equation is aquired using Ansys$^\text{\textregistered{}}$ Mechanical, using the thermal conductivities measured above. Option 1 was chosen for the ohmic contact. Due to the symmetry of the chip, only half of the quadrupler is simulated as shown in Figure 3. In this analysis, the total power dissipation level is estimated to be 10 mW per anode. Moreover, only heat conduction is considered. Furthermore, all the non-contact surfaces of the block are assumed to be at room temperature, i.e. $23.0\celsius$. The quadrupler chip is designed [3] such that it is suspended in the waveguide block, with protruding gold beam leads clamping it to the waveguide block at the input and output (see Figure 3(a)). Two additional beamleads protrude from hairpin bias chokes and are bonded to a quartz-supported filter that sits in the block. Figure 3(a) shows the steady state temperature distribution of the original quadrupler chip.  
%
%

\begin{figure}[h]
\centering
\includegraphics[width=3.5 in]{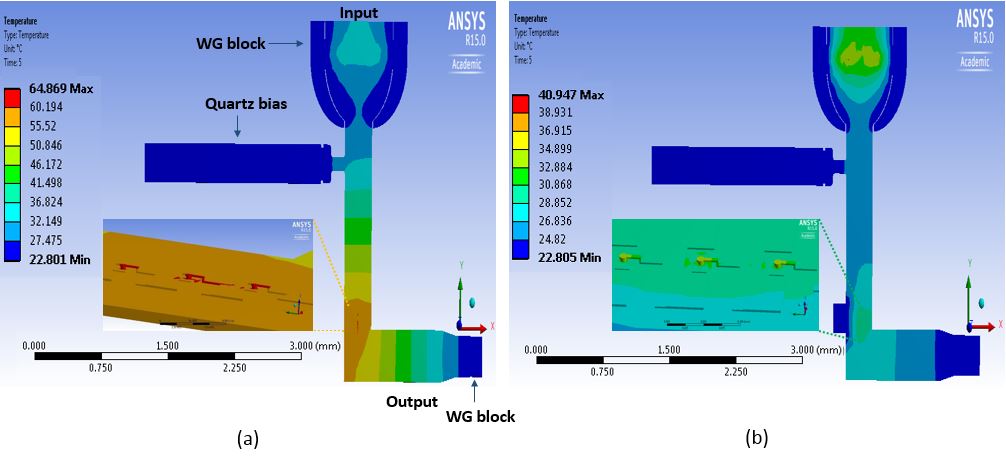}
\caption{Steady-state temperature distribution of: (a) Original quadrupler design, (b) Modified design with an extra beam lead.}
\label{fig_scattering}
\end{figure}

The second stage doubler diodes (magnified in the toggle window) experience the maximum anode temperature of $64.9\celsius$, due to the lack of proximity of a block heat sink. Figure 3(b) shows the temperature distribution with an extra beam lead close to the second stage diodes. This additional contact to the block reduces the maximum temperature experienced by the diodes to $41.0\celsius$.

%




\section{Conclusion}
In this paper, a steady state thermal analysis of a 160 GHz balanced quadrupler was performed. Time-domain thermoreflectance was used to characterize the thermal conductivities of the materials used in this analysis. Finally, a modified design with an additional beam lead showed an improvement in terms of thermal management.





%

\end{document}